\documentclass[iop]{emulateapj}
\usepackage{amsmath,amssymb}
\usepackage{multirow}
\usepackage{xcolor}
\usepackage{url}
\usepackage{natbib}
\usepackage[colorlinks,citecolor=blue,linkcolor=blue,urlcolor=black]{hyperref}
\usepackage{etoolbox}

\makeatletter

\patchcmd{\NAT@citex}
 {\@citea\NAT@hyper@{%
 \NAT@nmfmt{\NAT@nm}%
 \hyper@natlinkbreak{\NAT@aysep\NAT@spacechar}{\@citeb\@extra@b@citeb}%
 \NAT@date}}
 {\@citea\NAT@nmfmt{\NAT@nm}%
 \NAT@aysep\NAT@spacechar\NAT@hyper@{\NAT@date}}{}{}

\patchcmd{\NAT@citex}
 {\@citea\NAT@hyper@{%
 \NAT@nmfmt{\NAT@nm}%
 \hyper@natlinkbreak{\NAT@spacechar\NAT@@open\if*#1*\else#1\NAT@spacechar\fi}%
 {\@citeb\@extra@b@citeb}%
 \NAT@date}}
 {\@citea\NAT@nmfmt{\NAT@nm}%
 \NAT@spacechar\NAT@@open\if*#1*\else#1\NAT@spacechar\fi\NAT@hyper@{\NAT@date}}
 {}{}

\makeatother

\begin{document} 

\newcommand{\Rm} {${\cal R}^-(Z)\,$}
\newcommand{\Rp} {${\cal R}^+(Z)\,$}
\newcommand{\Q} {{ ??}}
\newcommand{\metal}{{\rm log(O/H)}+12 }
\newcommand{\Msun}{{\rm M}_\odot}

\newcommand{\specialcell}[2][c]{%
 \begin{tabular}[#1]{@{}c@{}}#2\end{tabular}}

\title{The Relative Rate of LGRB Formation as a Function of Metallicity}

\author{J. F. Graham$^{\hyperref[jfg]{1}}$\affil{Max-Planck Institute for Extraterrestrial Physics, Giessenbachstrasse 1, 85748 Garching, Germany}}
\author{A. S. Fruchter$^{\hyperref[af]{2}}$\affil{Space Telescope Science Institute, 3700 San Martin Drive, Baltimore MD 21218}}
\affil{$^1$\label{jfg} Max-Planck Institute for Extraterrestrial Physics, Giessenbachstrasse 1, 85748 Garching, Germany \\ $^2$\label{af} Space Telescope Science Institute, 3700 San Martin Drive, Baltimore MD 21218}

\journalinfo{}
\submitted{}

\begin{abstract}

There is now strong evidence that Long-duration Gamma-Ray Bursts (LGRBs) are preferentially formed in low-metallicity environments. However, the magnitude of this effect, and its functional dependence on metallicity have not been well characterized. In our previous paper, \cite{stats_paper}, we compared the metallicity distribution of LGRB host galaxies to the that of star forming galaxies in the local universe. Here we build upon this work by in effect dividing one distribution by the other, and thus directly determine the relative rate of LGRB formation as a function of metallicity in the low-redshift universe. We find a dramatic cutoff in LGRB formation above a metallicity of $ {\rm log(O/H)}+12 \approx 8.3 $ in the KK04 scale, with LGRBs forming between ten and fifty times more frequently per unit star-formation below this cutoff than above. Furthermore, our data suggests that the rate of LGRB formation per unit star formation continues to fall above this break. We estimate the LGRB formation rate per unit star formation may drop by as much as a factor of one hundred between one-third solar and solar metallicity.

\end{abstract}

\section{Introduction}

Shortly after Long-soft Gamma-Ray Bursts (LGRBs) were identified as extragalactic events spanning cosmological distances, it became apparent that they predominantly occur in blue, highly starforming and often irregular galaxies \citep{Fruchter1999, Fruchter, LeFlochblue, LeFlochblue2002, Christensen, LeFloch2006}. To determine whether this distribution of hosts was different from that expected from a sample of galaxies drawn randomly according to their rate of massive star formation, \cite{Fruchter} compared the hosts of LGRBs with those of Core-Collapse Supernovae (CCSNe) found in the Great Observatories Origins Deep Survey (GOODS). They found that while half of the GOODs CCSNe occurred in grand design spirals (with the other half in irregulars), only one out of the 18 LGRB host galaxies of a comparable redshift distribution was in a grand design spiral. Using an enlarged sample, \cite{Svensson} found a very similar result. Massive stellar progenitors should be just as available per unit star-formation in spirals as they are in irregulars, as it appears the stellar IMFs of blue irregulars and spirals are largely similar \citep{Bastian}, However, due to the the mass-metallicity relationship of galaxies \citep{Tremonti2004}, blue irregulars are typically far less-metal rich than spirals. This lead \cite{Fruchter} to conclude that LGRB formation is much more likely in low-metallicity environments.

A similar conclusion was reached by \cite{Stanek2007}, who showed that the very nearest LGRB hosts all have low metallicity when compared to similar magnitude galaxies in the Sloan Digital Sky Survey (SDSS) sample. Furthermore, \cite{Kewley2007} found the LGRB host sample to be comparable to extremely metal-poor galaxies in luminosity-metallicity relation, star-formation rate (SFR), and internal extinction. 

\cite{Modjaz2008} dramatically strengthened this result by taking advantage of the fact that a broad-lined Type Ic (Ic-bl) supernova has been found underlying the light of nearly every LGRB in which a deep spectroscopic search was performed (\citealt{Cano2014,Thone2014}).  \cite{Modjaz2008} showed that Ic-bl SNe with associated LGRBs are observed to occur in host galaxies with much lower metallicities than either the hosts of Type Ic-bl SNe without associated LGRBs or the bulk of the star-forming galaxies in the SDSS. This dramatic metallicity difference between the Ic-bl and LGRB samples suggests a metallicity dependent step in either the formation of the gamma-ray jet or in its ability to escape the progenitor which has either burned or lost its outer hydrogen and helium layers \citep{Langer}.

More recently, however, \cite{MannucciLGRBs} has suggested that the metallicity aversion of LGRBs is not intrinsic to their formation, but rather a consequence of a fundamental relationship between the mass, metallicity, and star-formation rates of galaxies \citep{Mannucci}. In this relationship the metallicity of a galaxy of a given stellar mass is anti-correlated with its SFR. Thus \cite{MannucciLGRBs} argued that the LGRB hosts are low-metallicity because they are effectively selected based on the basis of their higher then average star-formation. However this argument does not explain why LGRBs should preferentially choose irregular hosts more frequently than the general population of core collapse SNe \citep{Fruchter} or why the Type Ic-bl SNe without associated LGRBs do not show a preference for low-metallicity hosts comparable to the Type Ic-bl SNe associated with LGRBs (c.f. \citealt{Modjaz2008}).

In our preceding work, \cite{stats_paper}, we compared the metallicity distribution of the hosts of LGRBs with that of the hosts of several similar indicators of star-formation: LGRBs, Type Ic-bl, and Type II SNe.  We found that three quarters of the LGRB hosts have metallicities below 12+log(O/H) $<$ 8.6 { in the KK04 metallicity scale \citep{KobulnickyKewley}}, while less than a tenth of local star-formation is at similarly low metallicities. However, our supernova samples were statistically consistent with the metallicity distribution of the general galaxy population. Furthermore, we were able to show that as all the LGRBs in our sample are at redshifts lower than one, the general decrease of galaxy metalicities with redshift is far too weak an effect to account for the observed metallicity difference. The fact that LGRBs nearly always are associated with a Type Ic-bl SNe would suggest that LGRB progenitors probably have similar masses to those of regular Type Ic-bl, thus largely eliminating the possibility that the observed LGRB metallicity bias is somehow a byproduct of a difference in the initial stellar mass functions. Rather, metallicity below half-solar must be a fundamental component of the evolutionary process that separates LGRBs from the vast majority of Type Ic-bl SNe and from the bulk of local star-formation.

While this work shows that LGRBs exhibit a strong and apparently intrinsic preference for low metallicity environments some exceptions to this trend do exist --- three of the 14 LGRB in the sample possess abundances of about solar and above. These exceptions show that is it still possible to form an LGRB in a high metallicity environment albeit with greater rarity. If we wish, for instance, to use LGRBs to trace the star-formation of the Universe (cf. \citealt{Yuksel}) we must understand the conditions required for their production and thus the selection effects that could substantially bias our estimates. The implications of these high metallicity bursts are important not only for understanding the formation of LGRBs but also for our being able to use them as cosmological probes.

Metallicity is certainly critical in LGRB formation, but it might not be the only environmental factor of relevance. An additional observation of \cite{Fruchter} was that LGRBs are far more likely to occur in the brightest regions of their hosts than if they simply traced the light of their hosts. However, the CCSNe followed their hosts blue light distribution. \cite{Kelly2008} showed that Type Ic-bl trace the blue light of their hosts far more like LGRBs than typical SNe. This suggests that LGRBs (and Type Ic-bl in general) are formed from very massive progenitors { which generally} do not have time to travel far from their birth sites before exploding, { Typical CC-SNe, on the other hand, come from a wider range of masses, and are by number heavily weighted towards lower initial masses}. Thus the shape of the IMF, and in particular, the relative rate of formation of the most massive stars is critical. It is thus quite interesting that the work of \cite{Hakobyan} suggests (but does not conclusively show) that the ratio of Ibc SN to Type II SNe is higher in galaxies with disturbed morphology than in undisturbed galaxies. The work of \cite{Anderson} shows that Ib SNe come from more massive progenitors than Type II SNe but somewhat less massive than Type Ic SNe. Thus the result of \cite{Hakobyan} could be due to a change in the IMF, though it should be pointed out this effect is roughly a factor of three, or an order of magnitude less than the metallicity effect we report here.
 Additionally, \cite{Kelly2014} have shown that in comparison to SDSS galaxies the hosts of Ic-bl SNe and LGRBs of a given stellar mass exhibit high stellar mass and star formation rate densities, and as well as high gas velocity dispersions. CCSNe hosts show no such preference. Furthermore, \cite{Kelly2014} find that this preference cannot be explained as a byproduct of a preference for low metallicity environments. While the surface brightness of LGRB hosts may be biased by the general increase in specific star-formation rate with redshift \cite{Noeske2007}, the hosts of Type Ic-bl sample are all at low redshift and thus would not be affected by this bias. Thus while metallicity may not be the only environmental factor to affect the rate of LGRB formation, it does appear to be the dominant factor.

Recently, \cite{perley_cutoff} analyzed a large sample of LGRB host galaxy photometry and came to the conclusion that the LGRB formation rate drops precipitously at metallicities above solar. This is very similar to the result of \cite{Wolf} who, using the original \cite{Fruchter} sample of hosts, argued that if the observed preference for dwarf hosts was produced by a sharp cutoff in the host metallicity distribution, that cutoff would be at about $12 + \rm{log}(\rm{O}/\rm{H}) \approx 8.7$. However, because neither group had spectroscopic metallicity measurements for their host galaxies, both groups used the standard mass-metallicity relationship for galaxies (\citealt{perley_cutoff} used \citealt{Zahid2014} whereas \citealt{Wolf} used \citealt{Erb2006}) to convert host photometry to metallicities. This is problematic, as LGRB hosts are systematically biased low in metallicity for a given galaxy mass \citep{Modjaz2008, Levesque051022, Levesque2, stats_paper}. Assuming that LGRB hosts follow the field mass-metallicity relationship causes one to substantially overestimate their metallicity and thus the value of any metallicity cutoff. 

Here, we build upon the results derived in \cite{stats_paper} to estimate the relative LGRB formation rate per unit star formation rate as a function of metallicity. In order to do this we must normalize the LGRB rate to the rate of underlying star formation across different metallicities. However, we show that this can be done as a straightforward extension of the work presented in \cite{stats_paper}. We combine our estimates of the LGRB and star formation rates as a function of metallicity to answer a fundamental question of this field: how much more likely is an LGRB to form at one metallicity as compared with another? 
 We address the implications of this metallicity induced rate difference on the absolute LGRB formation rate in \cite{form_rate_letter}.

\section{The Metallicity Dependence of the LGRB Formation Rate}

\begin{figure*}[t]
\begin{center}
\includegraphics[width=.49\textwidth]{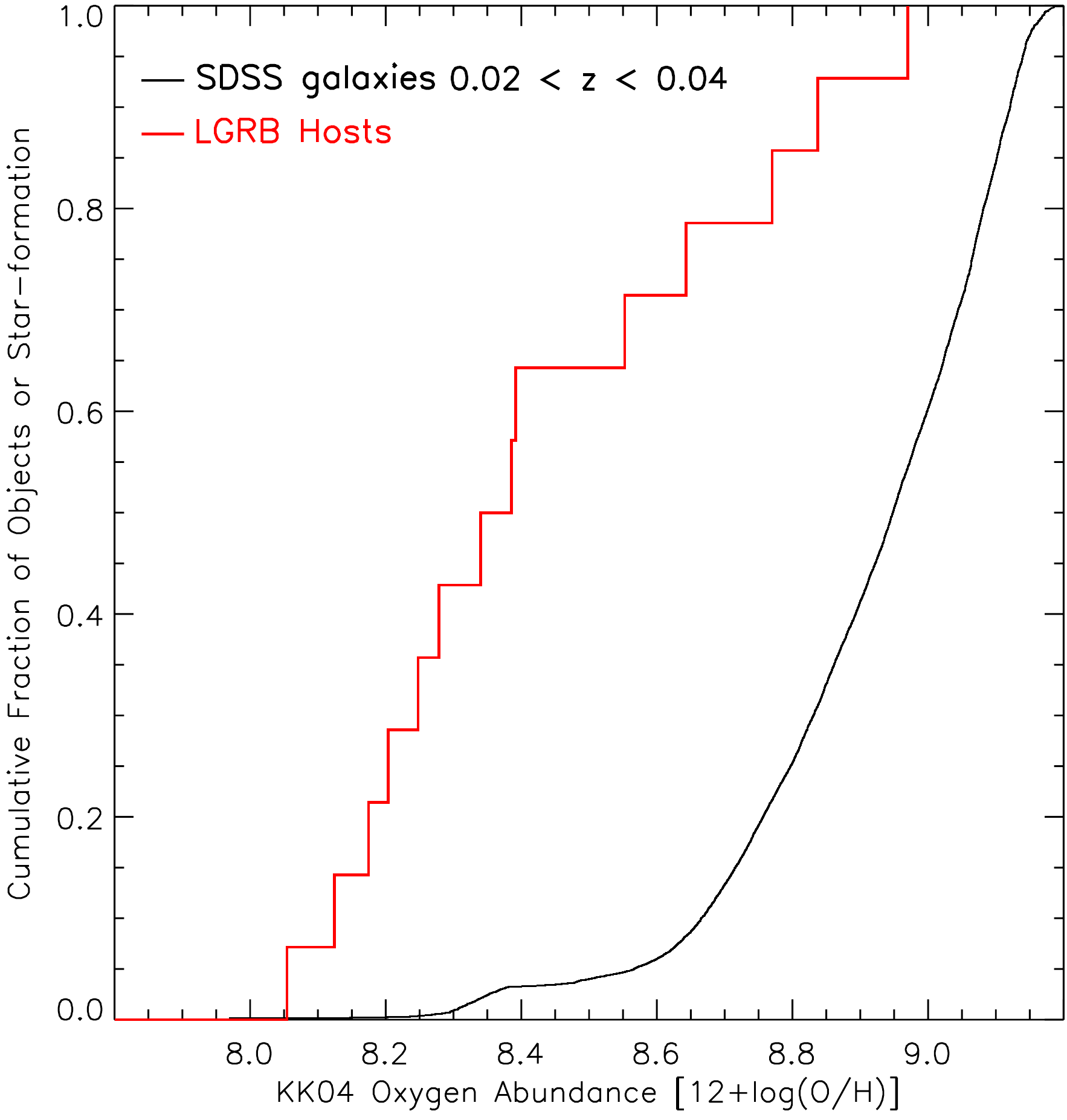}
\includegraphics[width=.496\textwidth]{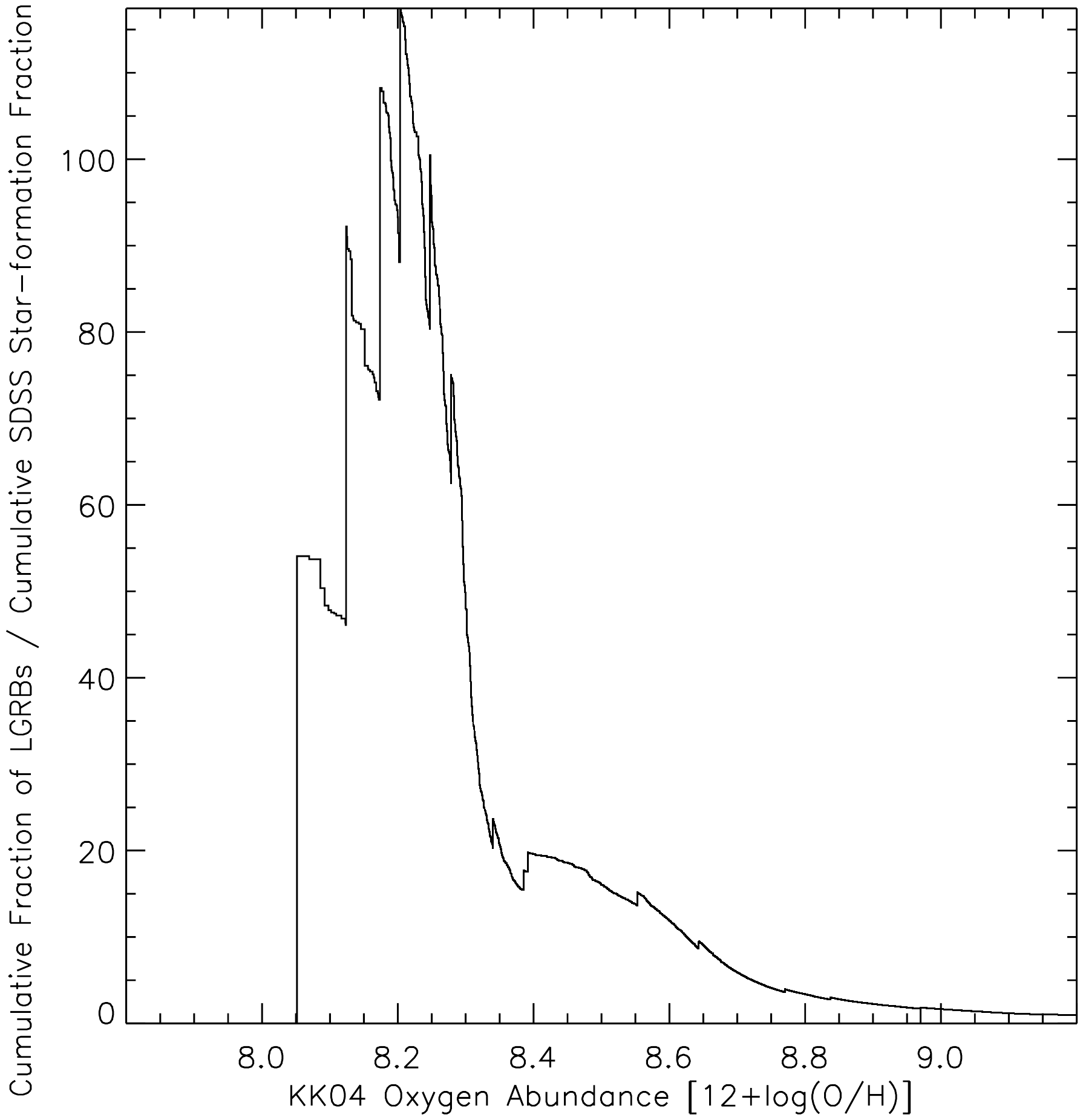}
\caption{\label{divide} Left: Cumulative fraction of population or total star-formation vs.\thinspace \thinspace galaxy central metallicity (see \citealt{stats_paper} Figure 5). The LGRB sample from \cite{stats_paper} is shown in red. The black line shows the local sample of SDSS star-forming galaxies also from \cite{stats_paper}. Each galaxy contributes according to its star-formation rate, as measured by its H$_\alpha$ emission. Right: Cumulative fraction of LGRBs divided by the cumulative SDSS star-formation fraction, or ${\cal R}^-(Z)$ as defined in Equation 1. For each $Z$, we show the average rate of LGRB formation per unit star formation for all $Z' \le Z$ normalized by the average rate of LGRB formation per unit star formation over the entire metallicity span. Note the sharp cutoff in the LGRB formation rate at about log(O/H)+12 $\sim $8.3 or at about 40\% of solar metallicity.}
\end{center}
\end{figure*}

In this section we estimate the relative rate of LGRB formation as a function of metallicity. To do this, we use the LGRB sample along with the comparison sample of Sloan Digital Sky Survey (SDSS) \citep{SDSSdr7,SDSS-mpg} star-forming galaxies used in \cite{stats_paper}. The metallicity distributions of these samples are shown in the left hand side of Figure~\ref{divide}. 
{ 
We adopt the LGRB sample of our prior work over the new and larger sample of \cite{xshooter_survey} because we have already considered potential biases on the LGRB population (see \citealt{stats_paper} section 2.4). We will briefly discuss possible selection effects in the following sections.

Our LGRB sample used all LGRBs with published spectra or line lists available (at the time) with the lines sufficient to obtain an R$_{23}$ metallicity with express [N II]/Halpha degeneracy breaking. { All metallicities that we present here are determined using the R$_{23}$ diagnostic in the KK04 scale, employing the iterative scheme of \cite{kd2002}.} The SDSS galaxy sample comprises spectroscopy of photometric catalogue galaxies complete down to an r-band Petrosian apparent magnitudes of 17.77 \citep{Strauss2002} to which we have applied a 0.02 $<$ z $<$ 0.04 redshift cut providing a complete volume limited sample of galaxies brighter than -18 B-band absolute magnitude. }

Whereas each LGRB host in Figure~\ref{divide} contributes equally to the cumulative LGRB host distribution, each SDSS galaxy contributes by its amount of star formation (estimated from its H$_\alpha$ emission, see \citealt{stats_paper}). This sample of SDSS galaxies comprises all the SDSS star-forming galaxies in the redshift range 0.02 $<$ z $\le$z 0.04 
with M$_B \le -18$ with spectroscopy suitable for metallicity measurement. The lower bound of this redshift range is necessary for the 3727 {\AA} [O II] line to enter the SDSS spectroscopic wavelength coverage, thus allowing us to apply the R$_{23}$ metallicity diagnostic (see \citealt{KobulnickyKewley} for a description of the diagnostic, code, and the KK04 scale). At the upper bound, the SDSS spectroscopy is complete to M$_B$ $\le$ -18 with partial coverage of dimmer galaxies. Twelve of the fourteen LGRB hosts in our sample are brighter than M$_B$ = -18, so this SDSS sample provides a good comparison. In the next section of this paper, we will discuss estimating an extension of the Sloan sample two magnitudes fainter, to provide an even better match to the LGRB sample. For an in depth discussion of the sample choice and the spectroscopic methods used in this paper, please see \citealt{stats_paper}, and in particular section 2.1 therein.

To estimate the relative rate of LGRB formation per unit star formation as a function of metallicity, we first introduce a bit of formalism. Define $f_{GRB}(Z)$ to be the LGRB formation rate per co-moving volume at metallicity $Z$ normalized so that $ \int_{-\infty}^\infty f_{LGRB}(Z)\, dZ = 1 $ and similarly let $f_{SFR}(Z)$ be the fractional star-formation rate at metallicity $Z$ per co-moving volume again such that $ \int_{-\infty}^\infty f_{SFR}(Z)\, dZ = 1 $. We then define 
\begin{equation}
{\cal R}^-(Z) = \frac{\int_{-\infty}^Z f_{GRB}(Z')\, dZ'}{\int_{-\infty}^Z f_{SFR}(Z')\, dZ'}
\end{equation}
where ${\cal R}^-(Z)$ is the ratio of these fractional rates up to a given metallicity $Z$, and the superscript $-$ on the ${\cal R}$ indicates we are taking a ratio from $-\infty$ to $Z$. Later, we will use a $+$ superscript to indicate the same ratio with the integrals are taken from $Z$ to $\infty$. The cumulative plots in Figure \ref{divide} are in fact plots of the numerator and denominator of this function. We plot the function ${\cal R}^-(Z) $ in the right hand side of Figure~\ref{divide}. For any metallicity $Z$, this function is the star-formation normalized LGRB rate below $Z$. This plot shows that the star-formation normalized LGRB rate is very high at low metallicities. This rate plunges at about $12 + \rm{log}(\rm{O}/\rm{H}) \approx 8.3$. By definition ${\cal R}^-(Z)$ converges to 1.0 as $Z \Rightarrow \infty$. At low metallicities, then, using this rough comparison, we appear to be seeing rates of LGRB formation at low metallicities perhaps fifty times greater than the average over the entire range of metallicity.

We call the present comparison rough because while we went to great lengths in \cite{stats_paper}, to produce a magnitude limited sample of SDSS galaxies for comparison, we used whatever LGRB galaxies for which we could obtain R$_{23}$ metallicities. This means that our results could be subject somewhat to the vagaries of our LGRB sample. In the following two sections, we will attempt to adjust our LGRB sample for potential biases, and will show that while the shape and the magnitude of ${\cal R}^-(Z)$ change in detail, its overall general properties appear to be largely independent of how we weight our LGRB host galaxies, or the exact comparison sample used. The methods we develop here will also be useful for application to the larger samples that are now becoming available. 

\subsection{The Effect of Dark Bursts}

Inclusion in this sample requires (high-energy) detection of the burst, localization of the burst to a clearly associated host galaxy, and then suitable spectroscopy of the host galaxy. Fortunately LGRBs themselves are detected by gamma-ray instruments which are not believed to be affected by the properties of their hosts. For example, \cite{Levesque_No_Correlation} finds no correlation between host metallicity and burst luminosity. Burst localization is however complicated by a subset of bursts which lack the optical counterpart usually used for determining the burst position to sub-arcsecond levels. Fortuitously the nature of these bursts, called ``dark bursts'', was of considerable interest when our LGRB sample was compiled and thus they received substantial attention. 

Dark bursts are now thought to simply be typical LGRB events obscured by dust in the host galaxy. They comprise twenty to perhaps up to thirty percent of of LGRBs \citep{Cenko_dark,Perley_dark_frac,Greiner2011}. Dust extinguished LGRBs are found to preferentially reside in more massive galaxies, which under the mass metallicity relation are typically more metal rich \citep{Kruehler2011, Perley_dusty_LGRBs}. We have 2 dark bursts (LGRBs 020819B and 051022) in our sample of 14 objects which is roughly consistent with the lower end of this rate. Our two dark bursts are among the three highest metallicity LGRBs in our sample. Thus if the sample were to contain something closer to $\approx$30\% dark bursts, we would have four dark bursts, and five rather than three very high metallicity hosts in the sample. Such a sample would produce results that were noticably, but not dramatically, different from those we will derive directly from the \cite{stats_paper} sample. However, it is generally straightforward to estimate the how the results would change with a higher proportion of dark (metal-rich) bursts, and we do so throughout the paper.

\subsection{Luminosity Completeness Adjustment of the Comparison Sample}

\begin{figure*}[t]
\begin{center}
\includegraphics[width=.50\textwidth]{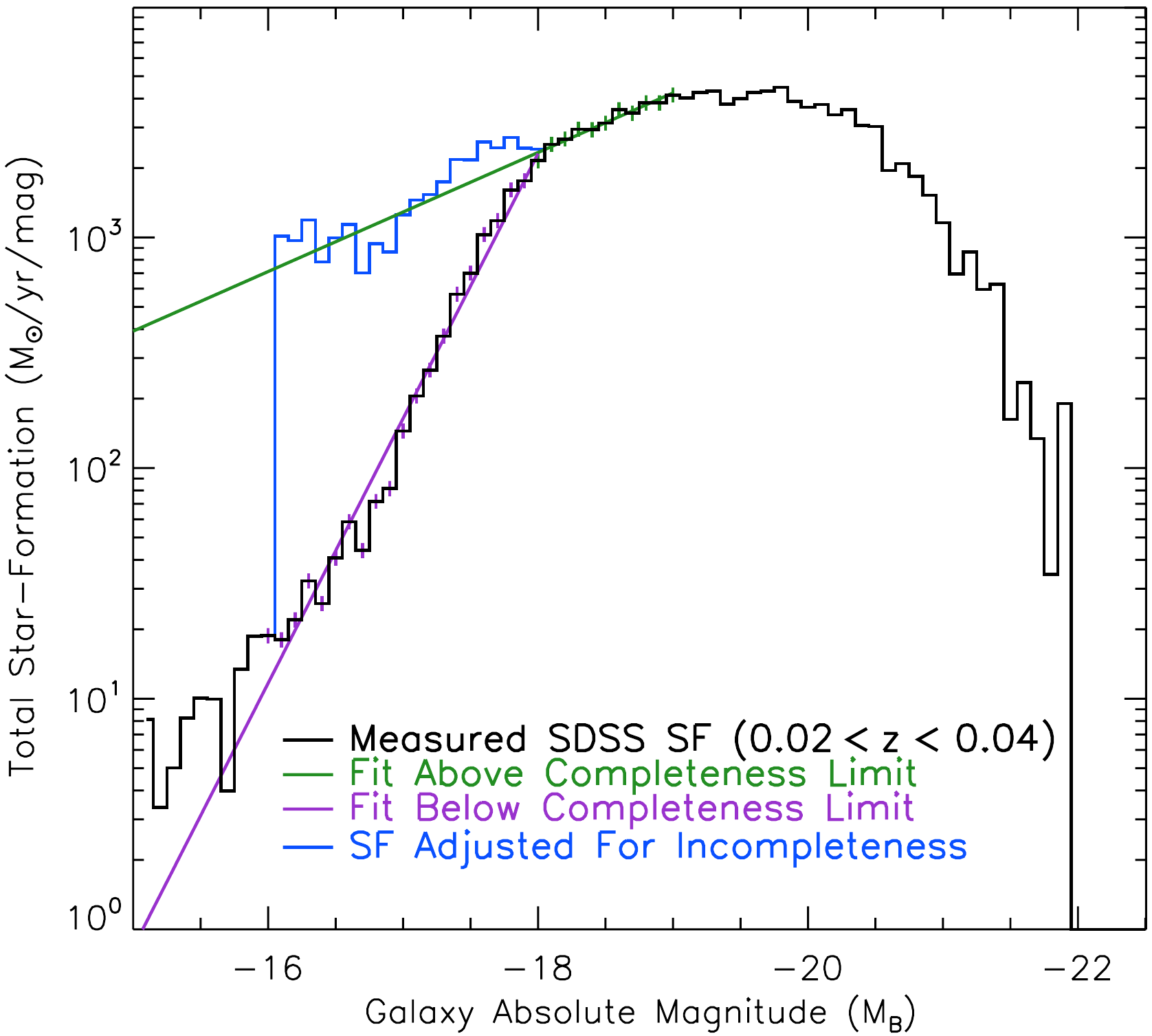}
\includegraphics[width=.43\textwidth]{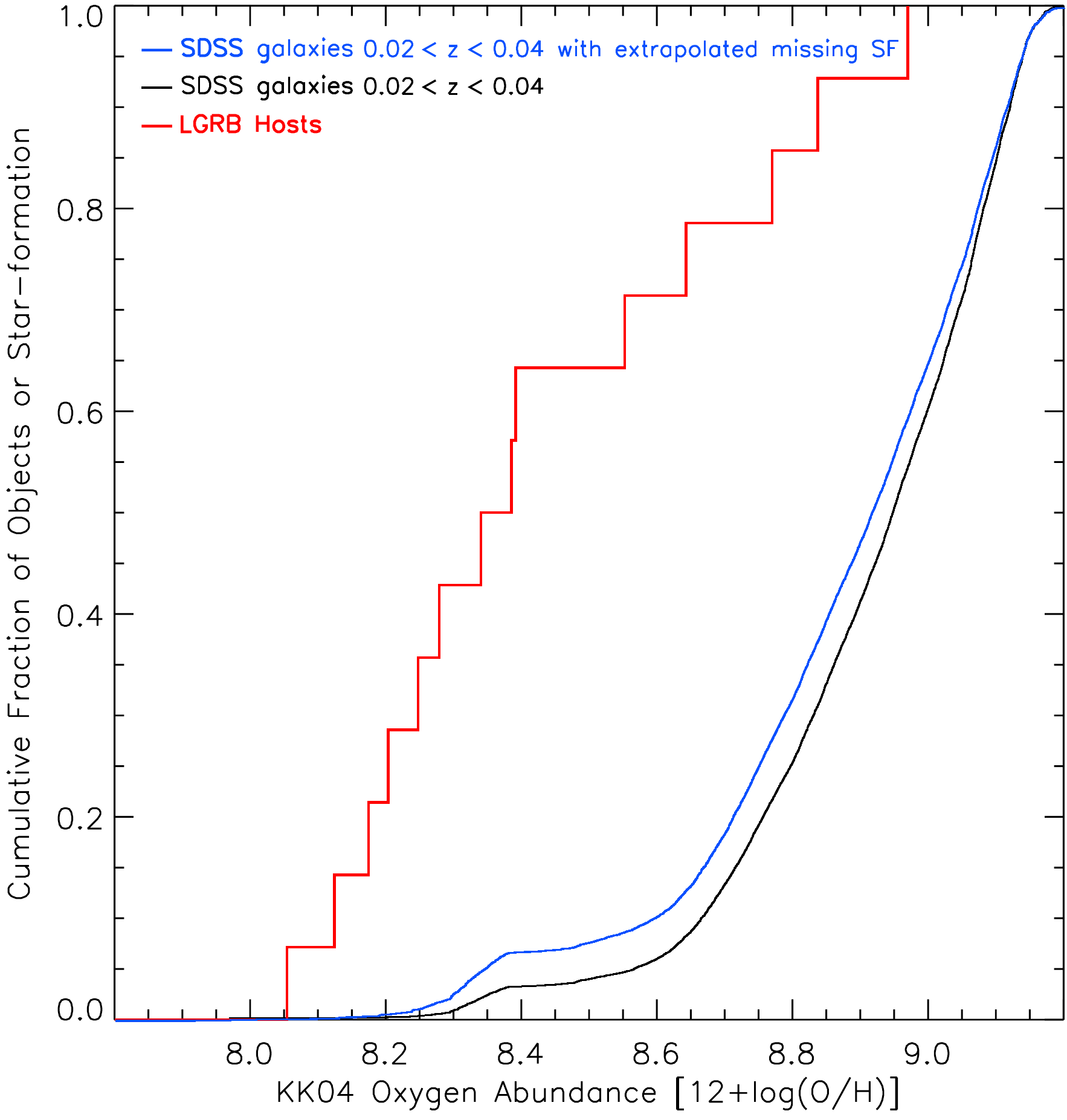}
\caption{\label{extrapolate} Left: Histogram of star formation in the SDSS sample. The black histogram shows the star formation per unit absolute magnitude in the volume limited SDSS comparison sample. This sample is complete only to M$_B$ -18. The purple line shows a fit to the raw sample at absolute magnitudes fainter than M$_B$ -18. The green line shows the power-law of the Schechter fit to this star-forming sample (see text of this paper and \citealt{stats_paper} for more details). We use the ratio of the Schechter power-law to the raw fit to boost the estimate the star-formation per unit magnitude down to M$_B$ -16 (blue line). Right: Cumulative fraction of LGRB population or total star-formation vs.\thinspace \thinspace galaxy central metallicity. The red line shows the LGRB metallicity distribution already seen in Figure \ref{divide} left. Similarly, the black line shows the distribution of star-formation versus metallicity for the SDSS sample down to our completeness limit of M$_B$ = -18, as in Figure \ref{divide} left. The blue line shows the star-formation distribution when the sample is extended to M$_B$ = -16.  Thus the blue and black lines represent the raw and extended star-forming sample respectively in both sides of the figure.}
\end{center}
\end{figure*}

The LGRB hosts in the present sample are as faint as M$_B \sim -16$, while our SDSS comparison sample { is only complete} down to M$_B =$ -18. Thus if we want to take full advantage of our LGRB sample, we need to correct for the incompleteness of our SDSS sample in the range $-16 < {\rm M}_B < -18$. As discussed in \cite{stats_paper}, the star-forming sample can be fit to a Schechter luminosity function with a power-law $\alpha = -1.3$. It is this power-law which determines the luminosity function at the faint end of the distribution. This power law is shown as the green line in Figure \ref{extrapolate} left. { Note how the slope of the histogram changes sharply for galaxies fainter than our completeness limit of M$_B =$ -18. To correct this, a} fit to the actual measured star formation in the incomplete sample at magnitudes fainter than M$_B = -18$ is shown as the purple line in that figure. To extend the sample below M$_B = -18$, we multiply the star-formation of any galaxy in the sample in that magnitude range by the ratio of the green (Schecter) fit to the purple (actual) fit at the galaxy's magnitude. That is we are boosting the star-formation in each galaxy in the sample between $-16 < {\rm M}_B < -18$ by the ratio of the shortfall of galaxies at that magnitude compared to the power-law extension of the luminosity function. The extended sample of galaxies can be seen as the blue continuation to the SDSS sample in Figure \ref{extrapolate} right. As each galaxy has a measured metallicity, the amount of star formation it contributes to its metallicity bin is also increased by the ratio of the two power laws. In Figure \ref{extrapolate} right we show the effect of this correction on the metallicity distribution of the SDSS star-formation. In particular the amount of star formation below a metallicity of log(O/H)+12 $ = $8.3 is roughly doubled. { Nonetheless, the SDSS star formation curve remains dramatically below that of the LGRBs at low metallicities.}

{ K-S tests confirm what is obvious to the eye in Figure \ref{extrapolate} right. Comparing the LGRB and original SDSS star-formation distributions we find a K-S value of 0.70 and a probability of 6.3 $\times$ 10$^{-7}$. Instead comparing the LGRBs to our extrapolated SDSS star-formation distribution gives a K-S value of 0.66 and a probability of 4.3 $\times$ 10$^{-6}$. Thus the LGRBs clearly exhibit a preference for lower metallicity environments. This low metallicity preference grossly exceeds the metallicity gradients available within galaxies. In \cite{stats_paper} we found that SNe host galaxies have the same metallicity distribution as the SDSS star-formation and estimate the metallicity difference between the SNe site locations and the galaxy centers. The average difference of 0.11 dex we would expect to also be roughly consistent with the correction between the central galaxy metallicities as measured by the SDSS fibers and the average metallicity of the star-formation in these galaxies. Even shifting the SDSS star-formation distribution metallicities down by three times this (which also happened to be the largest single expected gradient in the SNe sample) still gives a K-S value of 0.49 and a probability of only 1.6 $\times$ 10$^{-3}$. While a difference between central galaxy and GRB site metallicity would also be expected, the LGRBs are much more closely distributed on the brightest regions of their typically much smaller host galaxies \citep{Fruchter} and thus this effect would be reduced.

We also note the existence of a surprisingly flat region in the SDSS star-formation distribution between approximately 12+log(O/H) of 8.37 to 8.5 suggesting a sparsity of star-formation within this metallicity range. We believe this effect is actually caused by measurement error on the R$_{23}$ value. The range where this is observed (i.e. 8.37-8.5) is right at the intersection between the upper and lower branches of the R$_{23}$ diagnostic where a small change in R$_{23}$ reflects a large change in the metallicity (see \citealt{KobulnickyKewley} Figure 7). This scatter is observed in other works using the R$_{23}$ diagnostics on large samples (e.g. \citealt{Modjaz2008} Figures 5 \& 6). As this effect is equivalently present in both the SDSS and LGRB samples it should not effect the analysis which depends on the ratio of these two samples.}

\begin{figure*}[t]
\begin{center}
\includegraphics[width=.49\textwidth]{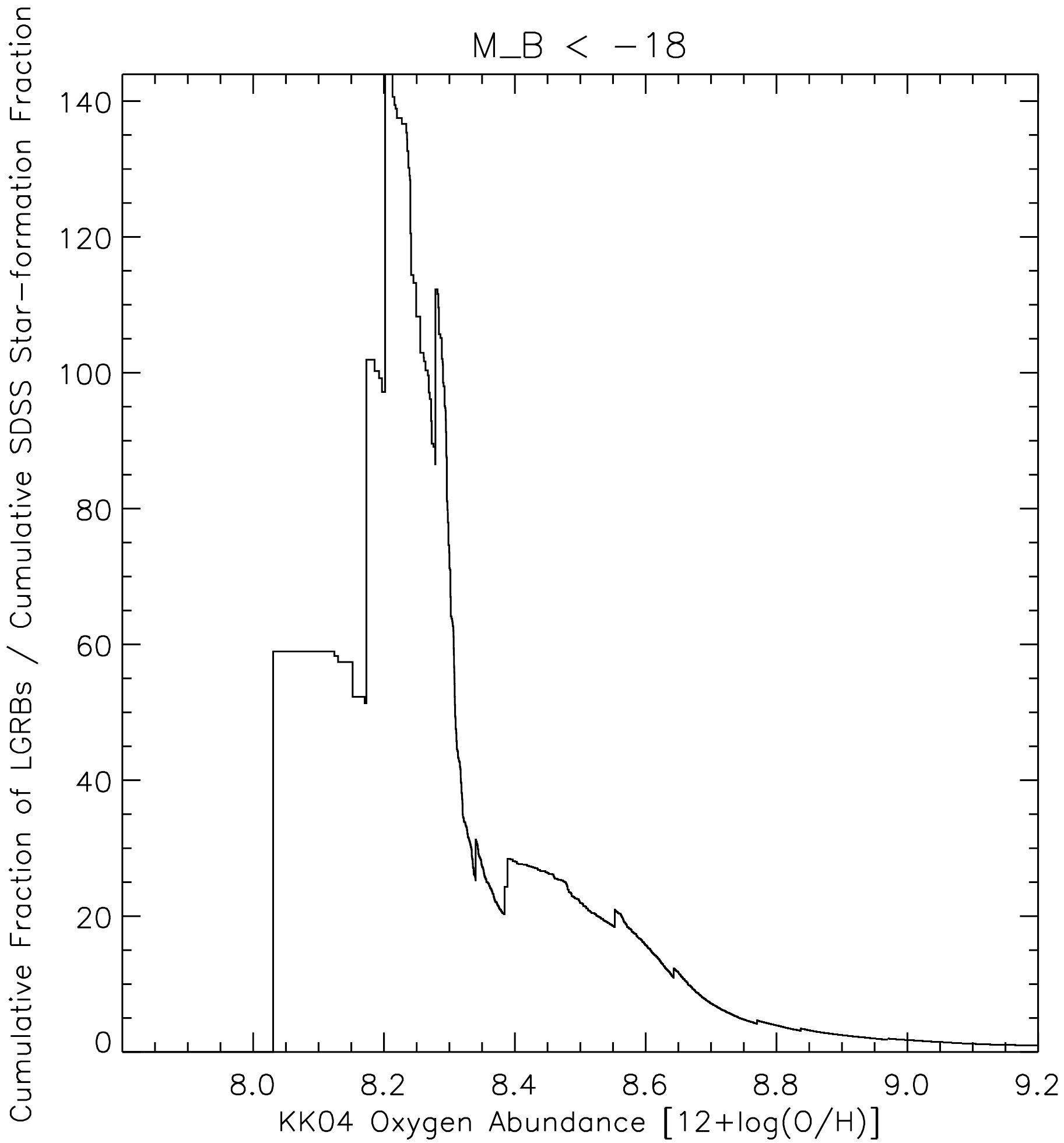}
\includegraphics[width=.481\textwidth]{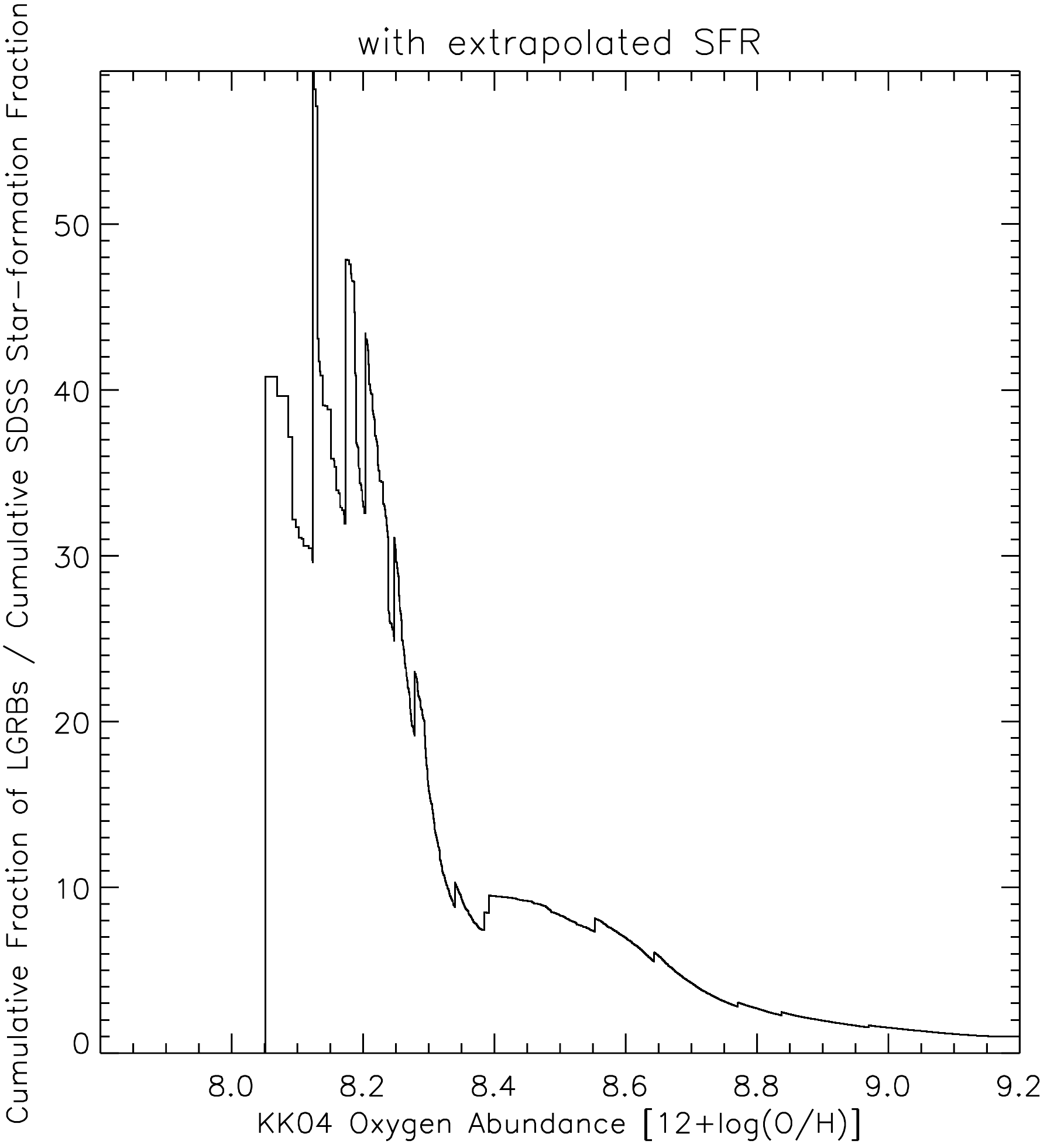}
\caption{\label{divide_more} Left: $\cal{R}^-(Z)$ for a magnitude limited sample. Here we show $\cal{R}^-(Z)$ where we restrict the LGRBs to those with hosts brighter than M$_B = -18$. In this case we do not need to use the extended SDSS sample, as the magnitude limits of the two sample are the same, but we must exclude two of our LGRBs. Right: The extended sample. Here we plot $\cal{R}^-(Z)$ for the full LGRB sample compared to the SDSS extended down to M$_B = -16$. Note the cutoff in the LGRB formation rate apparent in Figure \ref{divide} is still present in both versions of this figure, though the exact value of $\cal{R}^-(Z)$ at any particular metallicity does vary somewhat depending upon the sample used.}
\end{center}
\end{figure*}

 { We now use the original and extended SDSS star formation distributions} to recompute the cumulative fraction of LGRBs versus star formation as a function of metallicity or $\cal{R}^-(Z)$. In Figure \ref{divide_more} left we show $\cal{R}^-(Z)$ where we only consider those LGRB hosts brighter than M$_B = -18$. In this way, the LGRB hosts have the same magnitude limit as the SDSS star-formation sample, and the LGRB hosts will not be biased to lower metallicity due to a luminosity metallicity correlation. In Figure \ref{divide_more} right, instead of cropping the LGRB host sample we use our extension of the SDSS star-formation distribution down to M$_B = -16$ in the computation of $\cal{R}^-(Z)$. Again while the details of the curve vary, the same overall shape is revealed: ${\cal{R}}^-(Z) > 30$ for $Z < 8.3$, while falling sharply above that metallicity. Interestingly the rapid decline slows substantially as the metallicity increases further, and the decline in relative LGRB formation with increasing metallicity appears to be rather minor above a metallicity of about log(O/H)+12 $ = $8.7, a point we shall return to later.

\subsection{LGRB Host Observably Adjustments}

The distance to which an LGRB host can be observed and its metallicity measured will depend on its brightness, and in particular on the strength of its emission lines. As our LGRB sample was limited by our ability to get host metallicites, which requires bright emission lines, the sample is to first order a magnitude limited sample and thus more luminous hosts will be overrepresented. However, we can adjust for this effect on ${\cal R}^-(Z)$ by weighting the contribution of each LGRB inversely by the volume over which it could have been discovered. 

 To calculate the distances over which a host might be detected, we estimate the luminosities of the 3727 {\AA} [O II], 4959 {\AA} H$\beta$, 5007 {\AA} [O III] lines\footnote{Other lines used in the metallicity diagnostics are not similarly considered for the following reasons: the 4959 {\AA} [O III] line has $\frac{1}{3}$ the 5007 {\AA} lines flux as quantum mechanically required (see \citealt{070714Bpaper}) thus only the brighter line is needed. The Balmer decrement sets the H$\alpha$ line at 2.87 times the lines H$\beta$ flux as required by Case B recombination (see \citealt{Balmer_decrement}) with the value increasing with extinction, and the 6584 {\AA} [N II] line is highly dependent on metallicity such that it is typical excluded from S/N cuts to avoid introducing a metallicity bias (see \citealt{stats_paper}).}, and for the least luminous line determine the redshift (and corresponding comoving volume) where it would be detected with a flux of 1$\times$10$^{-17}$ erg sec$^{-1}$ cm$^{-2}$ (assuming standard cosmological parameters: $\Omega _m$ = 0.27, $\Omega _\lambda$ = 0.73, \& H$_\circ$ = 0.71 km s$^{-1}$ Mpc$^{-1}$). An emission line with this flux would have about a five sigma detection in a four hour integration on an eight-meter class telescope. \label{z_limit_description} 

\begin{table*}[t]
\begin{center}
\caption{\label{GRB_table} LGRB Sample.}
\vspace{-0.1 cm}
\begin{tabular}{lcccccccccccccccccccccc}
\hline
\hline
\multirow{2}{*}{~LGRB host} & \multirow{2}{*}{Metallicity} & \multirow{2}{*}{M$_B$} & \multirow{2}{*}{z (observed)} & \multirow{2}{*}{z (limit)} & CMV (at limit) & CMV (capped) \\
 & & & & & Mpc$^3$ & Mpc$^3$ \\
\hline
GRB 991208 & 8.05 & -18.68 & 0.706 & 1.31 & 273 & 158 \\
GRB 030329 & 8.12 & -16.52 & 0.1685 & 0.94 & 132 & 132 \\
GRB 070612A & 8.17 & -20.86 & 0.671 & 3.44 & 1360 & 158 \\
GRB 011121 & 8.20 & -19.75 & 0.362 & 2.91 & 1080 & 158\\
GRB 060218 & 8.24 & -15.92 & 0.034 & 0.44 & 19.9 & 19.9 \\
GRB 031203 & 8.27 & -18.52 & 0.1055 & 1.42 & 321 & 158\\
GRB 010921 & 8.34 & -19.87 & 0.451 & 1.29 & 264 & 158\\
GRB 020903 & 8.38 & -19.34 & 0.251 & 1.40 & 312 & 158\\
GRB 050824 & 8.39 & -19.02 & 0.828 & 1.43 & 325 & 158 \\ 
GRB 980425 & 8.55 & -18.09 & 0.0085 & 0.72 & 70.6 & 70.6\\
GRB 060505 & 8.64 & -19.38 & 0.0889 & 1.18 & 219 & 158\\
GRB 051022 & 8.77 & -21.23 & 0.80625 & 3.24 & 1250 & 158 \\
GRB 050826 & 8.83 & -20.28 & 0.296 & 1.40 & 312 & 158\\
GRB 020819B & 8.97 & -21.53 & 0.411 & 1.25 & 247 & 158\\
\hline
\end{tabular}
\end{center}
\vspace{-0.2 cm}
In this table we present the LGRB hosts in our sample, sorted by redshift. We report the host M$_B$, observed redshift and the maximum redshift at which all of the lines necessary for determining the metallicity could be detected with a signal to noise ratio of at least five (See Section \ref{table_discussed}). The CoMoving Volumes (CMV) are given for both the limiting redshift and the redshift capped at z = 1.
\end{table*}

\begin{figure}[h!]
\begin{center}
\includegraphics[width=.48\textwidth]{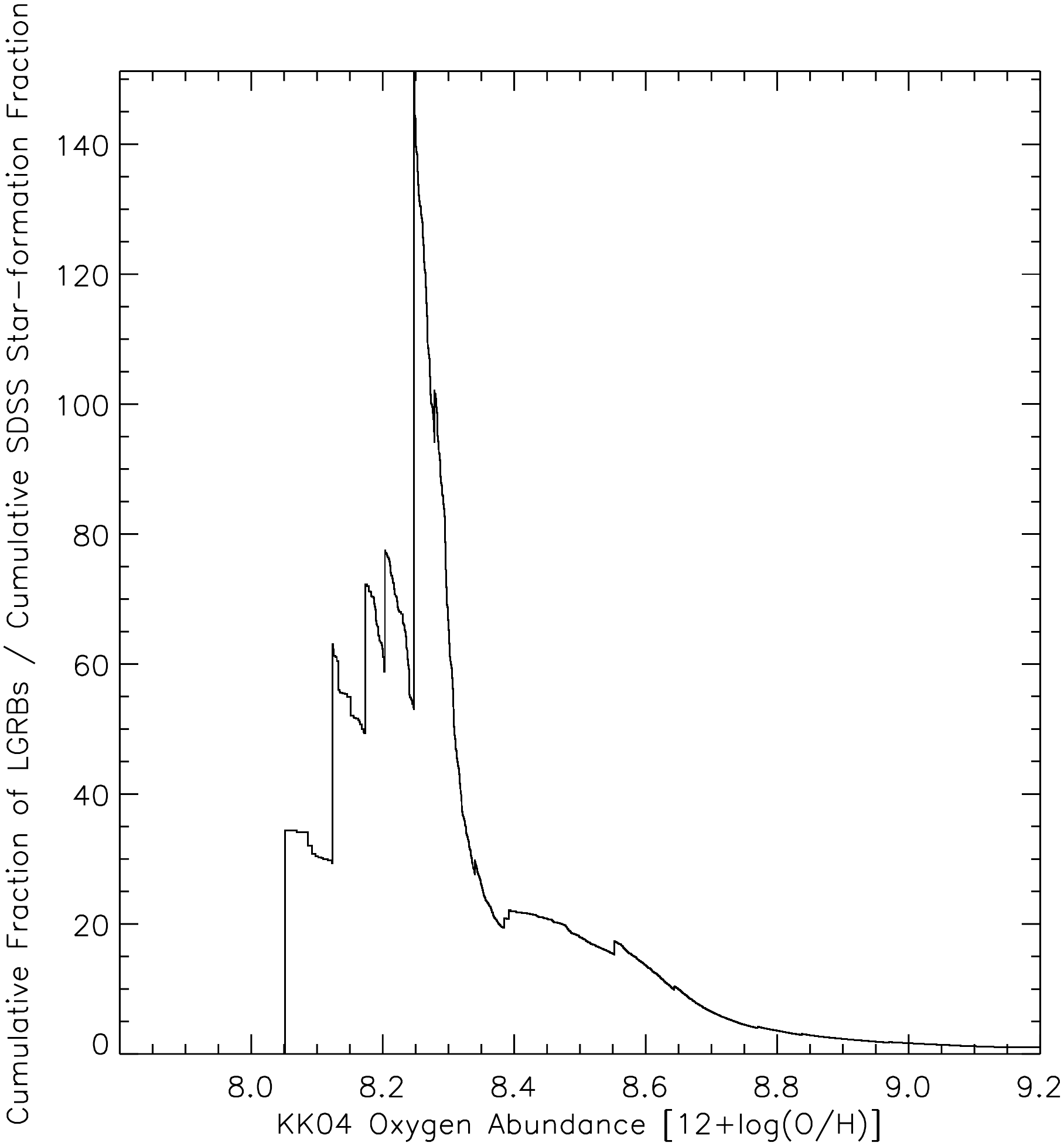}
\caption{\label{LGRB_CMV}\label{LGRB_num_stat}\Rm weighted by CMV. In this plot, LGRBs are weighted by the inverse of the CoMoving Volume (CMV) when calculating \Rm out to the redshift to which their hosts' metallicities could have been measured. The CMV is capped using the lesser of a redshift of $z=1$ or the maximum redshift at which the faintest host line necessary for determining the metallicity would be expected to have a flux above 1$\times$10$^{-17}$ erg sec$^{-1}$ cm$^{-2}$. The hard limit of $z=1$ is a function of 
the surveys we used in compiling the sample and the need for deep spectroscopy in the near-IR at redshifts above one. More recent surveys now coming available will allow an extension of the techniques we describe here to higher redshift.}

\end{center}
\end{figure}

However, the surveys we have used gave either line equivalent widths or line fluxes uncorrected for slit loss.  For the two closest objects (LGRBs 980425 and 060505) we use star-formation rates from the literature (\citealt{Christensen980425} and \citealt{Ofek} respectively) to estimate the total H$\alpha$ flux and then rescale the other lines accordingly. The remaining objects are sufficiently distant that slit losses are not large, and we are able to estimate the slit losses as discussed in \citealt{stats_paper}. For objects where we have a measure of the H$\beta$ line equivalent width and B \& V band absolute magnitude values, we can estimate the H$\beta$ flux independently and scale the spectrum according. Although we do not present numbers from this method here, they are comparable to those obtained using our slit loss estimates. In many cases, the redshift out to which we estimate we could accurately determine the host metallicity is quite high. However, none of our objects are much above $z=0.8$. This is because of the surveys which were used to compile our sample, and the technical difficulties of determining galaxy metallicities above $z=1$. \label{table_discussed} While surveys are now becoming available that will be able to reach these higher redshifts, we estimate that our sample was effectively limited at $z=1$ and thus we cap the comoving volume associated with any host to the comoving volume out to $z=1$. In Table \ref{GRB_table}, we give the observed and limiting redshifts for the hosts as well as their estimated and capped co-moving volumes. The LGRB hosts are sorted by their redshift to allow easier identification of specific LGRBs with features in the plots. In Figure \ref{LGRB_CMV} we plot the relative rate of LGRB formation versus metallicity, \Rm, with LGRBs weighted inversely to their capped CVM. While the peak of \Rm just before the fall is higher in this plot than earlier ones, the general shape of the plot,
and the location of the sharp cutoff are essentially unchanged.

\section{Relative Rates of LGRB Production}

\begin{figure*}[t!]
\begin{center}
\includegraphics[width=.49\textwidth]{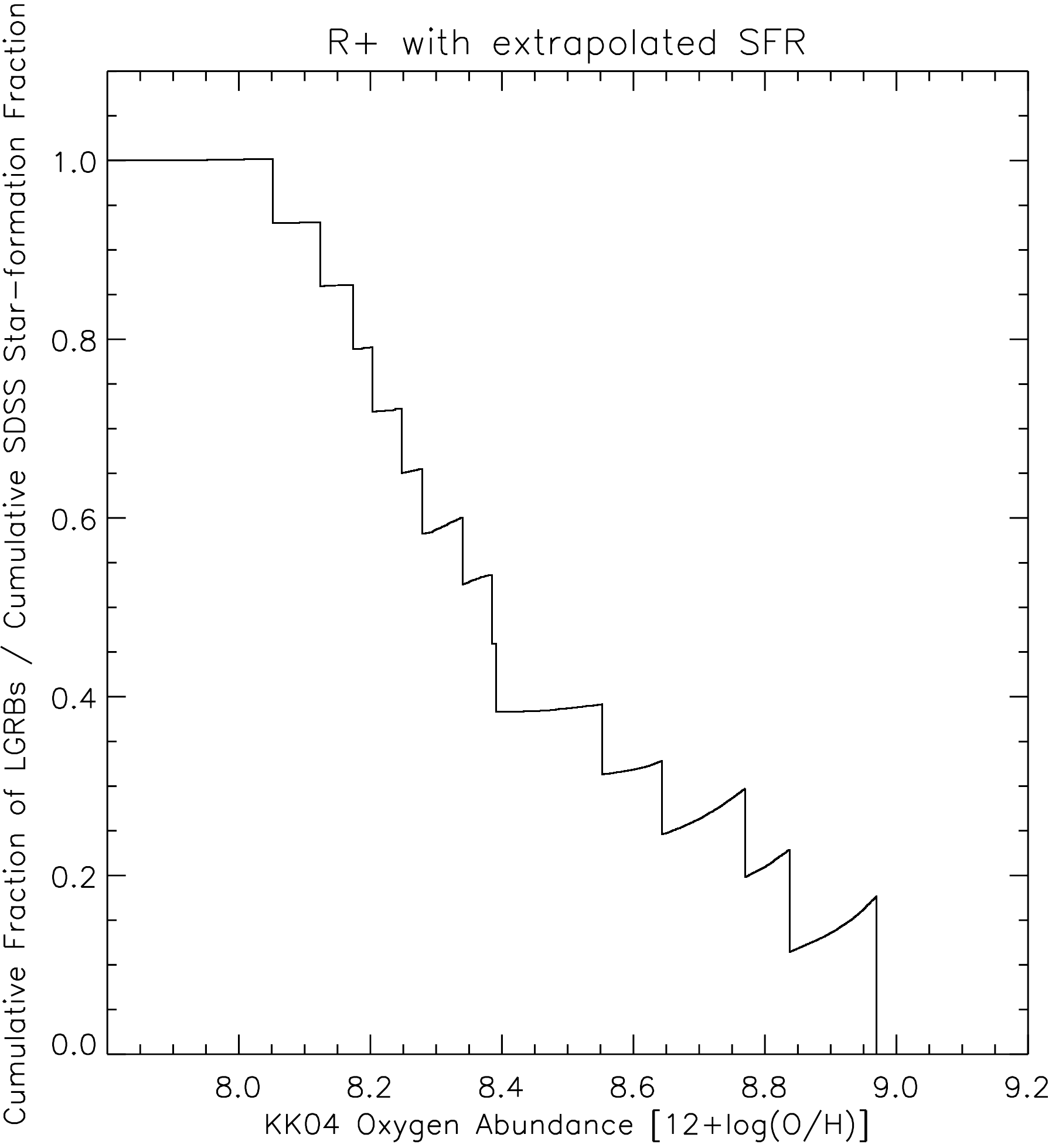}
\includegraphics[width=.49\textwidth]{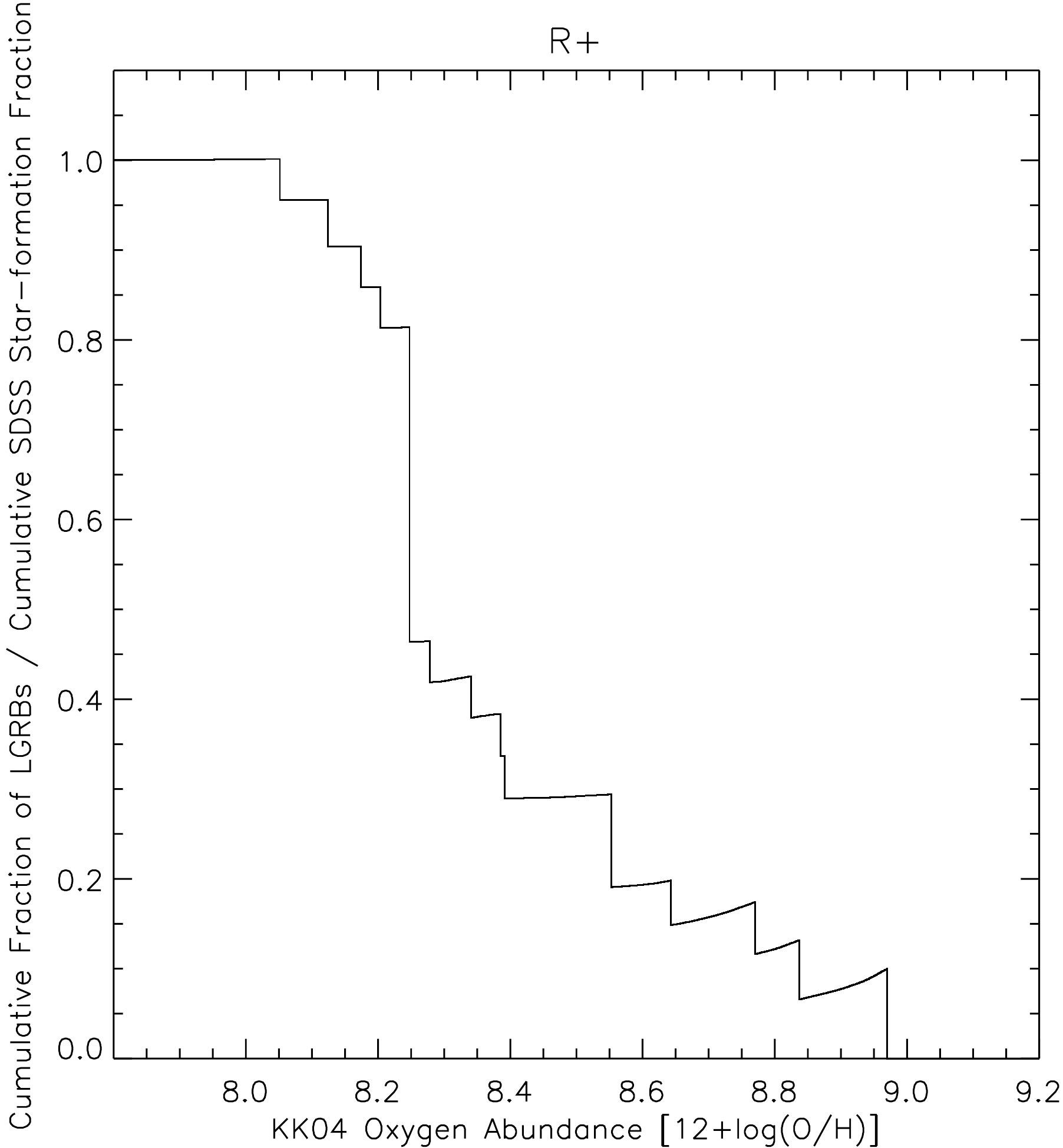}
\caption{\label{Figure_R+} The function $\cal{R}^+(Z)$. This function is described Eq.~\ref{R+}. It is the integral of the fraction of LGRBs in the sample over the integral of the fraction of total star formation, where both integrals are taken over the range $Z \rightarrow \infty$. Each step in the plot is a separate LGRB. In both cases shown here we use the SDSS sample extended to $\rm M_B =-16$ On the left we show the value of this function where each LGRB is weighted equally. On the right we show the value of this function where the LGRBs are weighted by the their comoving volume, capped by the voulume within the sphere out to $z = 1$, and where LGRBs above $z=0.5$ are weighted by an additional factor of 2.5 to take into account their apparent underrepresentation in our sample (see Figure~\ref{LGRB_CMV}). The curvature in the individual steps (which is particularly pronounced at higher values of $Z$) is caused by the integrated star formation between the discrete metallicities of the different LGRBs.}
\end{center}
\end{figure*}

So far we have entirely relied upon using the function ${\cal R}^-(Z)$ to represent the relative rate of formation of LGRBs as a function of metallicity. However, this is a somewhat blunt, if powerful, tool. By definition ${\cal R}^-(Z)$ converges to 1.0 as $Z \Rightarrow \infty$. Thus the rate of LGRB formation over a range of metallicity ${\infty, Z}$ is measured relative to the entire range, i.e.~${-\infty, \infty}$. To allow us to better compare rates in differing ranges of metallicity, we introduce ${\cal R}^+(Z)$, where 
\begin{equation}
\label{R+}
{\cal R}^+(Z) = \frac{\int_Z^{\infty} f_{GRB}(Z')\, dZ'}{\int_Z^{\infty} f_{SFR}(Z')\, dZ'}.
\end{equation}
 This is the normalized star-formation rate of LGRB production above metallicity $Z$. We plot ${\cal R}^+(Z)$ in Figure~\ref{Figure_R+}. 
Then if we define
\begin{equation}
{\cal R}^D(Z_1,Z_2) = \frac{{\cal R}^-(Z_1)}{{\cal R}^+(Z_2)} ,
\end{equation}
${\cal R}^D(Z_1,Z_2)$ gives us the star-formation normalized rate of LGRB formation below $Z_1$ divided by the star-formation normalized LGRB formation above a metallicity of $Z_2$. This allows us to directly compare the rate of LGRB formation at low metallicities to that at high metallicities.

When calculating these functions, we need to choose between the weighting schemes we have discussed in the previous sections. Our most conservative estimate of the LGRB formation rate at low metallicities is produced by comparing the complete sample of unweighted LGRBs to the extended star-formation sample (see Figure \ref{divide_more} right). A more aggressive approach is that shown in Figure \ref{LGRB_CMV} right, where the LGRBs are weighted by their (capped) co-moving volumes. Therefore for the rest of this section we will calculate values for both these approaches, with the difference between them giving some estimate of the importance of systematic effects in determining the result.

\begin{figure*}[t!]
\begin{center}
\includegraphics[width=.48\textwidth]{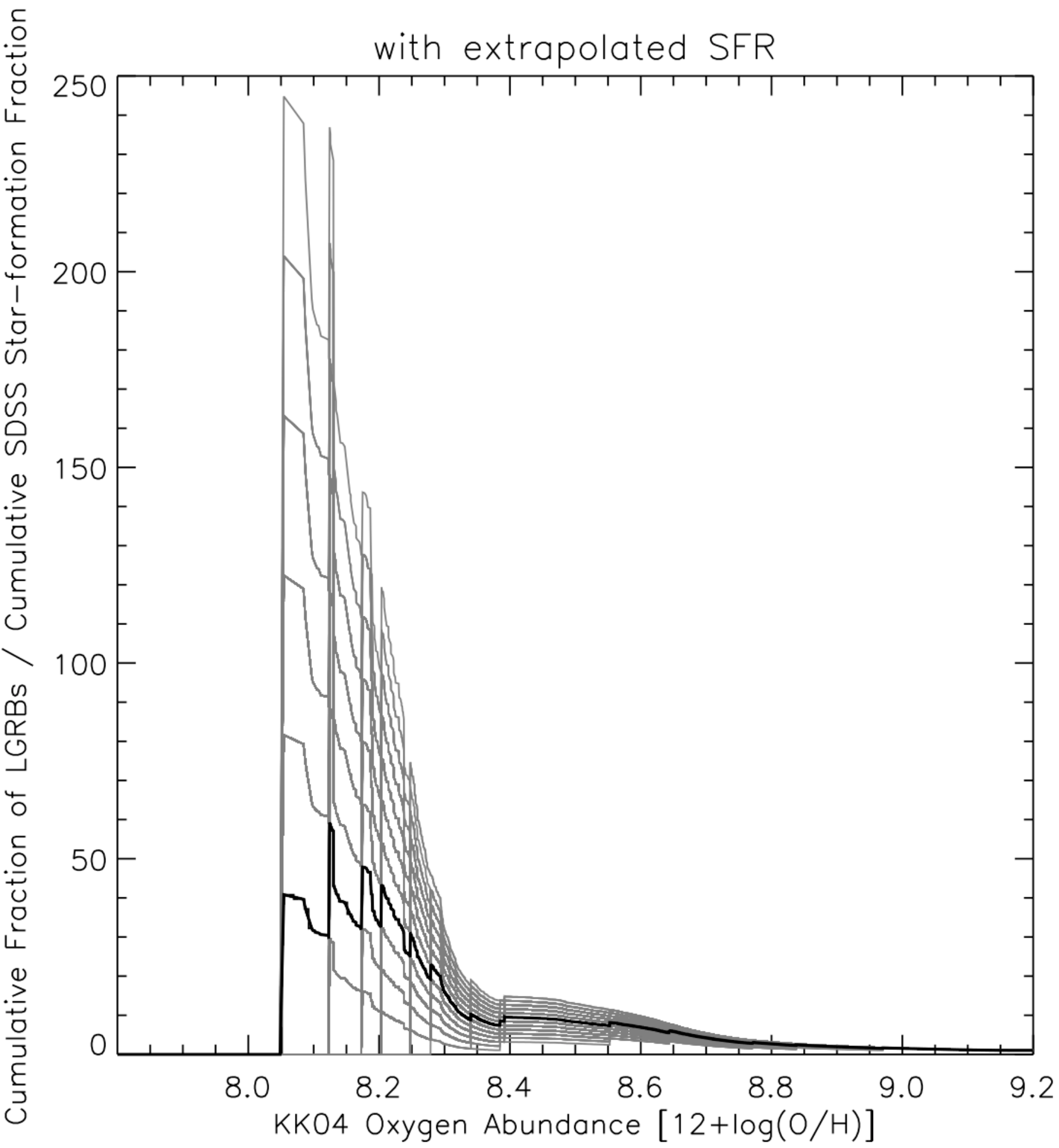}
\includegraphics[width=.48\textwidth]{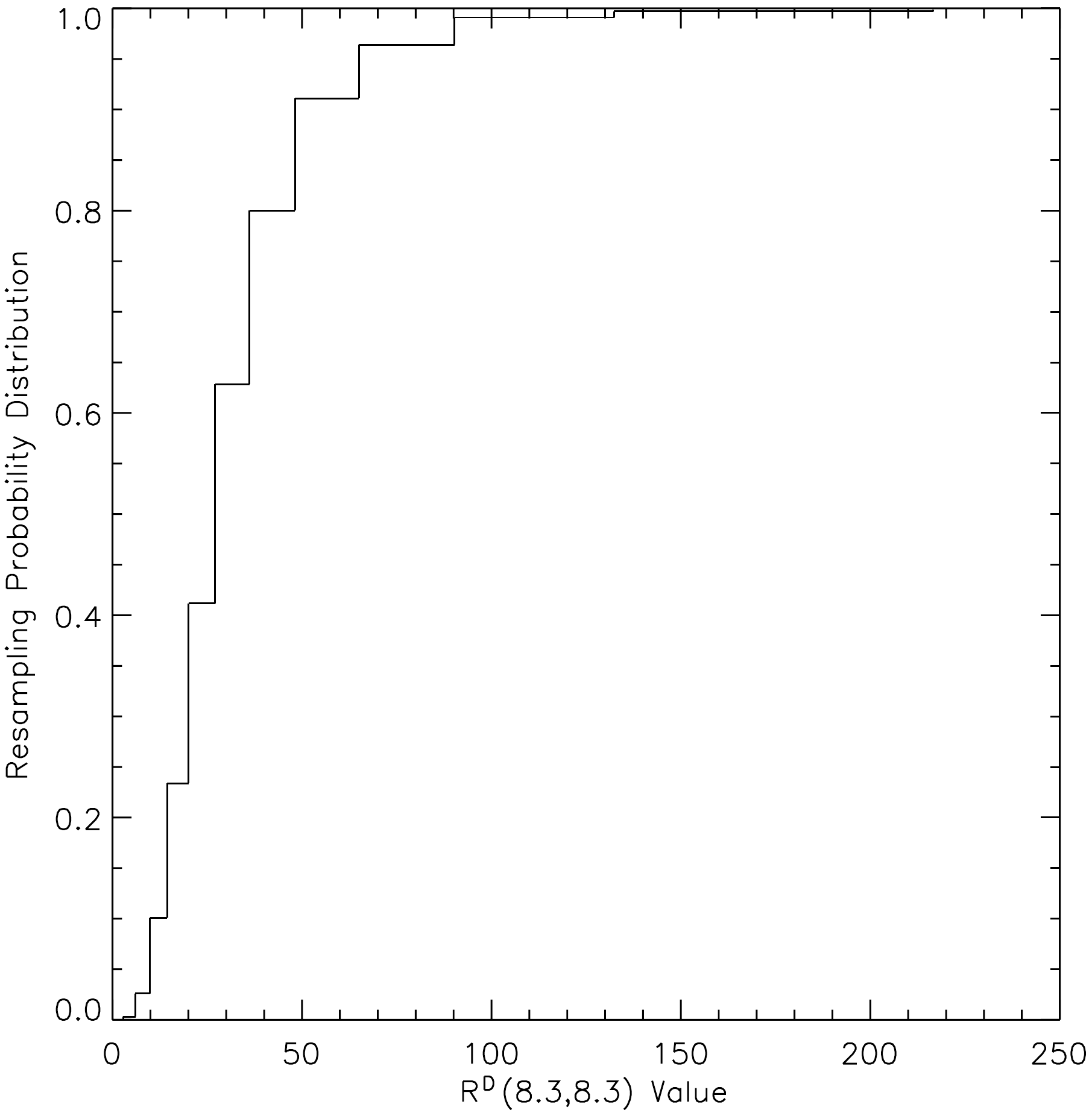}
\caption{\label{resample} { Left: $\cal{R}^-(Z)$ shown for 1000 iterations of re-sample with replacement using our LGRB population and the SDSS with cutoff of$\rm M_B =-18$. The dark line is $\cal{R}^-(Z)$ shows the result when the resampling returns our original sample. This line is in fact Figure \ref{divide_more} right.
Note that a significant dependence on metallicity remains for all iterations, though the strength of the effect varies. Right: A cumulative distribution of the ${\cal R}^D(8.3,8.3)$ values generated in the sample and replace. This effectively give us a cumulative probably distribution of the ${\cal R}^D(8.3,8.3)$ values. Based on the resampllng there is only a 2.5\% chance that ${\cal R}^D(8.3,8.3)$ is below a factor of ten, and over a 90\% chance that it is between 10 and 50. This is, however, likely a conservative estimate, as use of the CMV weighted sample causes the rate estimates to roughly double.}}
\end{center}
\end{figure*}

A clear division between low and high-formation rates is created by the rapid drop in ${\cal R}^-(Z)$ centered on 
${\rm log(O/H)} + 12 \approx 8.3$. If we then compute ${\cal R}^D(8.3,8.3)$, we obtain the relative rate of LGRB formation per unit star formation above 8.3 compared to that below 8.3. If we do not weight the LGRBs, and compare them to the SDSS star-formation rate extended down to an absolute magnitude $\rm M_B =-16$, i.e. our conservative case, we find ${\cal R}^D(8.3,8.3) \approx 27.1$. As there are seven LGRBs in the sample with metallicity below 8.3 and seven above that metallicity, the statistical errors in this estimate are large, about 40\%. If we instead use the LGRBs weighted by their (capped) co-moving volumes, we find ${\cal R}^D(8.3,8.3) \sim 150$. Here, however, one LGRB in particular, GRB 060218, makes a substantial contribution because of the relatively low comoving volume over which the metallicity of its host could be measured. { Removing this one object drops the estimate of ${\cal R}^D(8.3,8.3)$ down to $\sim$65 and we use this reduced LGRB sample in our subsequent CMV estimates.

We also examine the effects of our small LGRB sample size through the use of resample and replace. We recreated a 1000 GRB host metallicity samples by randomly choosing metallicities from our present population. In Figure \ref{resample} left we plot the resulting $\cal{R}^-(Z)$ distributions. The general shape of the distribution is largely invariant to the resampling, and in particular the sharp drop at log(O/H)+12 = 8.3 remains. On the right hand side of the same figure we show the cumulative distribution of values of ${\cal R}^D(8.3,8.3)$ under the resampling. We find a population mean of ${\cal R}^D(8.3,8.3)$ of 30.9, with ${\cal R}^D(8.3,8.3)$ lying between 10 and 50 in over 90\% of the trials. Use of the CMV weighted sample, on the other hand, causes the mean of and the limits on the ${\cal R}^D(8.3,8.3)$ distribution to go up by more than a factor of two. }

The effect of adjusting these numbers for our possibly low proportion of dark bursts is small in comparison to the statistical errors. If one assumes that instead of two dark bursts our sample should have four, and if one further assumes that these are high metallicity objects, and that they have weights on average equal to those of the high metallicity objects, then the numbers given above should be multiplied by a factor of $\sim 7/9$. Thus, we would find ${\cal R}^D(8.3,8.3) \sim 20$ in our conservative estimate, { and $\sim$50 for the co-moving volume weighted estimate.}

If we are willing to work with somewhat smaller statistics we can make a greater separation between the low and high regions of metallicity we use for comparison. The normalized rate of LGRB production is roughly constant (and high) for $Z < 8.2$ in the uniformly weighted sample. If we take the region $Z > 8.6$ for our upper range, we will be comparing to the LGRB production rate at roughly solar metallicity. We find that ${\cal R}^D(8.2,8.6) \sim 100$ for the uniformly weighted LGRBs, and ${\cal R}^D(8.2,8.6) \sim 300$ for the sample weighted by the capped CMVs. Here we only have four LGRBs in each of the two samples, and thus statistical errors will be in the range of 70\%. In this case, the effect of correcting for dark bursts would be somewhat larger, as we would add two additional bursts to the four in the high metallicity subset. Here the results would be multiplied by a factor of $\sim 2/3$, meaning ${\cal R}^D(8.2,8.6) \sim 66$ for the uniformly weighted sample, and $\sim 200$ for the sample weighted by capped CMVs.

\subsection{Correcting for Metallicity Evolution}
While galaxies in our SDSS sample have redshifts $0.02< z < 0.04$, the LGRB hosts in our sample go out to $z \sim 0.8$. However, the metallicities of galaxies of a given mass evolve with redshift, with the least massive galaxies showing the greatest change in metallicity with redshift. Indeed, at a redshift of $z \sim 0.8$ galaxies with a stellar mass of $3 \times 10^9$ M$_\odot$ have metallicities more than $0.15$~dex below that of galaxies of a similar mass at redshift zero \citep{Zahid2013}. This may seem like a small change, but in our extended SDSS sample the fraction of star formation with $ \metal < 8.2 $ is only one quarter that with metallicities $ \metal < 8.3 $. As a result, our estimates of relative rates of LGRB production could be significantly affected by ignoring this effect.

We do not know the distribution of star formation as a function of metallicity at higher redshifts. However, as a first approximation it is reasonable to estimate that the fraction of star formation below a metallicity $Z$ at a redshift of $z$ is equal to the fraction of star formation below a metallicity of $Z + \Delta_z$ at redshift zero, where $\Delta_z$ is the change in metallicity of galaxies between redshift $0$ and $z$. $\Delta_z$ depends both on the redshift under consideration and the mass of the galaxy. The smaller the galaxy mass, the larger the effect. For galaxies of mass of only $\sim 10^9 \, {\rm \Msun}$ at $z = 0.8$, \cite{Zahid2013} find $\Delta_z \sim 0.15$. However, there is a small difference between our sample and that of \cite{Zahid2013}: our sample only goes out to $z=0.04$ while theirs extends to $z=0.08$. Now, essentially all of the galaxies fainter than $\rm M_B = -18$ in the \cite{Zahid2013} sample are also in our sample (due to their limiting magnitude in the SDSS
falling inside $z=0.04$). However, our samples will differ somewhat for brighter galaxies. The more distant galaxies in \cite{Zahid2013} will be smaller on the SDSS spectroscopic fiber than equivalent ones in the nearer sample. The fiber will therefore obtain a central metallicity over a larger area of the galaxy, slightly reducing the estimated metallicity. We have compared galaxies brighter than $\rm M_B = -18$ in the SDSS out to the two different limiting redshifts, and find that this effect only produces a $\Delta_z \approx 0.05$. Again, this effect is only important for the brighter galaxies, where the metallicity evolution seen by \cite{Zahid2013} is quite small. However, we find that when taking these two effects into account for all the galaxies in our sample with metallicities below $8.3$, these galaxies all remain below roughly equal to 8.3. As the functions \Rm and \Rp do not depend on the distribution of LGRBs above or below a metallicity $Z$, but only the (weighted) fraction of LGRBs below or above that metallicity, the estimates of ${\cal R}^D(8.3,8.3)$ do not change. {\it Thus our estimates of ${\cal R}^D(8.3,8.3)$ are quite robust.}

Another quick way to check this calculation is to examine the metallicity of the five bursts with $z < 0.2$. These low redshift bursts would be expected to be subject to minimal metallicity evolution. Three of these five bursts are are below $\metal \sim 8.3$ even though less than three percent of the star formation in our extended sample is in this range. This small sample gives a value of ${\cal R}^D(8.3,8.3)$ about a factor of two larger than that reported for the entire sample. Although the statistical error is large, the fact that this estimate suggests an even greater bias in LGRB formation again supports our claim that our estimate of ${\cal R}^D(8.3,8.3)$ is robust.

However, we cannot make a similar claim for our estimates of ${\cal R}^D(8.2,8.6)$. This result could vary by up to a factor of four, depending how one attempts to take metallicity evolution into account. However a reduction of a factor of four brings ${\cal R}^D(8.2,8.6)$ down only to the level of ${\cal R}^D(8.3,8.3)$. As we would expect ${\cal R}^D(8.2,8.6)$ to be the larger of the two, this suggests that our estimates of the effect of metallicity on LGRB production are indeed quite robust, and that {\it at a very minimum} the rate of LGRB production increases by a factor of twenty-five higher between metallicities $ \metal < 8.3 $ and solar metallicity.

\section{Summary and Conclusions}

We have shown a dramatic cutoff in the rate of LGRB formation per unit star-formation at metallicites above log(O/H) + 12 $\approx$ 8.3, where our metallicity is determined using the KK04 scale of the R$_{23}$ metallicity diagnostic.  To test the stability of this result we have subjected our samples to a number of corrections intended to remove possible biases. 

First we address a known limitation of the SDSS survey, that the SDSS fiber placement is complete only for objects brighter than about 18th magnitude in the B band. This corresponds to a galactic luminosity of M$_B \sim$ -18 at z = 0.04, our highest volume limited redshift. The simplest correction is to just discard the two LGRB hosts and all the SDSS galaxies fainter than this absolute luminosity. We also employ a more detailed correction were we model the SFR as a function of galactic luminosity, extrapolate the missing star-formation, estimate how under surveyed the fainter galaxies in the SDSS are, and overweight them such that their total SFR matches our extrapolations. This allows us to preserve our estimates of star-formation as a function of metallicity down to a luminosity of -16 magnitude consistent with the luminosity range of our LGRB sample. Both of these changes leave the result intact.

Next we weight the hosts by the volume searched. Based on the LGRB hosts in the sample we conclude that the hosts would have remained in the sample so long as the key lines for determining metallicity were all brighter than 1$\times$10$^{-17}$ erg sec$^{-1}$ cm$^{-2}$. We then calculate the maximum redshift where the weakest line necessary for determining the host metallicity would thus be observable. We can then weight each object in our LGRB sample by the inverse of the comoving volume for its maximum observable redshift. However, as our sample relied upon line measurements obtained in the optical, no objects in the sample are at a redshift greater than one, and thus we limit the comoving volume to that of $z=1$. 

Additionally, our sample fourteen hosts contains two dark bursts, or a dark burst fraction of fourteen percent, somewhat below the estimated rate of dark bursts of twenty to perhaps thirty percent \citep{Cenko_dark,Perley_dark_frac}. Where we cite LGRB formation rate, we include a correction that would effectively bring our dark burst fraction up to thirty percent, yet this correction is in all cases is no larger than our statistical errors.

Throughout all of these correction{ s}, we continue to find a sharp cutoff at log(O/H) + 12 $\approx$ 8.3 in ${\cal R}^-(Z)$, the relative rate of LGRB formation below metallicity $Z$.
The dramatic cutoff is followed by a more gradual decline, which itself again appears to be greatly reduced above a metallicity of log(O/H) + 12 $\approx$ 8.7. However, the small number of LGRBs in our sample above this metallicity makes it difficult to accurately determine the slope of this fall-off. 
However, we 
note that this result agrees well with the claim of \cite{obs_paper} that the three highest metallicity objects in this sample are completely consistent with the mass metallicity relationships at their redshifts. Nor do these results necessarily conflict with the result of \cite{Greiner2015} that LGRB hosts appear to show little metallicity dependence at 3 $<$ z $<$ 5: presumably, by these redshifts, there would be relatively little star-formation above the metallicity cutoff.

Given the reasonably sharp nature of the observed cutoff observed at a metallicity of log(O/H) + 12 $\approx$ 8.3, we attempt to quantify how much more likely an LGRB is to form at metallicites below as opposed to above this cutoff. To do this, we determine the relative rate of LGRB formation below this cutoff divided by the relative rate of LGRB formation above the cutoff, or in new formalism, ${\cal R}^D(8.3,8.3)$. We find that this ratio is at least a factor of about twenty and quite possibly significantly larger, depending on which weighting scheme is used. We also see evidence that ${\cal R}^D(8.2,8.6)$, the ratio of the relative rate of LGRB formation below a metallicity of log(O/H) + 12 $\approx$ 8.2 to that above 8.6 may be close to a factor of one hundred. 

{ Indeed, the enhancement in the rate of LGRB production at low-metallicity is in fact so great that a substantial fraction of Type Ic-bl SNe at low metallicities may be required to supply LGRB production, a point made in detail in \cite{form_rate_letter}. 
However, it is also becoming clear that this extraordinary dependence on metallicity may not be restricted to LGRBs. Hydrogen poor superluminous supernovae (Type I SLSNe) also display a remarkable aversion to near solar metallicities. All of the Type I SLSNe of \cite{Chen2016} and all of the sample of Type I SLSNe in \cite{Perley2016} (with perhaps one exception) have metallicities ${\rm log(O/H)} + 12 < 8.5$. Now these samples were largely found in surveys that placed a significant emphasis on following up supernovae in faint (or non-visible) hosts. Thus this sample may exaggerate the dependence on metallicity, but it nonetheless seems highly likely that SLSNe share at least as strong an aversion to high metallicities as LGRBs. 

{ While we calculate our metallicities using the R$_{23}$ diagnostic in the KK04 scale, there are a number of alternate means of determining metallicity using strong emission lines  (e.g. \citealt{McGaugh, Z94, kd2002, D02, Tremonti2004, pp04, Nagao, PMC2009, Marino2013, MoralesLuis2014, Blanc2015, Dopita2016}). While these diagnostics vary in log(O/H) + 12 values by typically a few tenths of a dex, we do not believe our fundamental results will depend strongly on the choice of diagnostic used.  \cite{KewleyEllison} examined these differences and determined transformation equations between several of these commonly used diagnostics.  While there is some scatter, a galaxy with a low, medium, or high metallicity relative to the galaxy population in one diagnostic would still be expected to be at a low, medium, or high metallicity relative to the galaxy population in other diagnostics.  This should apply equally to host galaxies and the galaxies in our Sloan comparison sample.  Thus the preference of LGRBs to occur much more often per unit star-formation at low metallicities should remain consistent regardless of how the metallicity is measured.  Indeed, we have cross checked our results in the T04 scale of \cite{Tremonti2004} and found the basic conclusions of this paper to be unchanged (though the exact metallicity of the curoff, for instance, does change, as expected, with a differing choice of metallicity indicator).}  
 
Although the mechanism through which metallicity affects the formation of LGRBs is uncertain \citep{Langer, Podsi}, it is widely believed that LGRBs require high angular momenta to produce jets, and that low-metallicity is likely required to maintain angular momentum through the evolution of the progenitor. Interestingly, a leading model for the engine in SLSNe is rapidly rotating, highly magnetized neutron stars, or magnetars (c.f. \citealt{KasenBildsten}). Indeed, it has been suggested that magnetars may power both LGRBs and SLSNe \citep{Metzger2015}, and one ``ultra-long" GRB, is associated with an unusually bright SN \citep{GreinerNature}. GRB 111209A, whose prompt emission lasted for about fifteen thousand seconds, or more than a factor of a hundred longer than typical LGRBs, had an underlying supernova, SN 2011kl, with an absolute magnitude of about ${\rm M}_V = -20$, \citep{GreinerNature} or roughly a magnitude brighter than the SN associated with a typical LGRB, and a magnitude fainter than a typical SLSN. Whether a magnetar model can account for both LGRBs and SLSNe, or whether another process, such as accretion onto a remnant black hole, powers LGRBs, it seems likely that low metallicity plays an important role in the preservation of the crucial angular momentum in both of these rare, powerful explosions.

Metallicity is not the only significant environmental factor in the production of LGRBs. As \cite {Kelly2014} has reported, star formation density may also affect LGRB formation. This may be related to the enhancement in the rate of Type Ibc formation in galaxy mergers found by \cite{Hakobyan}. However, this effect only produces an enhancement by a factor of $\sim$3, not the factor of $\sim$30 we see in LGRBs due to the effect of metallicity Thus while many processes may contribute to the formation of LGRBs, our result strongly suggests that metallicity is the predominant determinant of the rate of LGRB formation relative to the rate of star formation.}

\acknowledgments

We thank Patricia Schady and Jochen Greiner for helpful comments. { We also thank Jarle Brinchmann for his help in determining the T04 metallicities of the hosts in our sample, which we used as a cross-check on the KK04 results presented here.}

John Graham acknowledges support through the Sofja Kovalevskaja Award to Patricia Schady from the Alexander von Humboldt Foundation of Germany.

\acknowledgments

\noindent {\bf Note added in proof:} \cite{Perley020819B} has shown that the metallicity value for GRB 020819B published by \cite{Levesque020819B} and used by us was in error.  This was due to a problem with the slit placement in the original observation.  The host of GRB 020819B has the highest metallicity of all the GRB hosts in our sample.  The high rates of LGRB formation at low metallicity which we derive here thus would have been even slightly higher had this been know earlier and had GRB 020819B been excluded.

\bibliographystyle{apj_links}
\bibliography{\jobname}

\end{document}